\documentclass{ifacconf}
\usepackage{subfig}
\usepackage{amssymb}
\usepackage{mathrsfs}
\usepackage{graphicx}      
\usepackage{algorithm}
\usepackage{algpseudocode}
\usepackage{amsmath}
\usepackage{natbib}        

\newtheorem{problem}{Problem}
\newtheorem{assumption}{Assumption}
\newtheorem{definition}{Definition}
\newtheorem{remark}{Remark}
\usepackage{url}
\begin{document}
\begin{frontmatter}

\title{Efficient Neural Hybrid System Learning and Transition System Abstraction for Dynamical Systems\thanksref{footnoteinfo}} 

\thanks[footnoteinfo]{This work was supported by the National Science Foundation, under NSF CAREER Award no. 2143351, NSF CNS Award no. 2223035, and NSF IIS Award no. 2331938.}

\author[First,Second]{Yejiang Yang}, 
\author[First]{Zihao Mo}, 
\author[First]{Weiming Xiang}

\address[First]{School of Computer and Cyber Sciences, Augusta University, Augusta GA 30912, USA (email: \{yeyang, zmo, wxiang\}@augusta.edu).}
\address[Second]{School of Electrical Engineering, Southwest Jiaotong University, Chengdu, China.}

\begin{abstract}               
This paper proposes a neural network hybrid modeling framework for dynamics learning to promote an interpretable, computationally efficient way of dynamics learning and system identification. First, a low-level model will be trained to learn the system dynamics, which utilizes multiple simple neural networks to approximate the local dynamics generated from data-driven partitions. Then, based on the low-level model, a high-level model will be trained to abstract the low-level neural hybrid system model into a transition system that allows Computational Tree Logic Verification to promote the model's ability with human interaction and verification efficiency. 
\end{abstract}


\end{frontmatter}

\section{Introduction}

In recent years, the development of neural networks has received particular attention in various fields, including natural language processing \cite{wang2023fusing}, computer vision \cite{stefenon2022classification}, etc. The applications of neural networks in system identification hold significant promise for they provide a precise approximation of the dynamics while requiring no prior knowledge of the system's mechanism. Neural networks serve as a predominant approach in machine learning, renowned for their exceptional ability to model complex phenomena with limited prior knowledge. Their proficiency in capturing intricate patterns in data offers valuable insights for dynamical system modeling, verification, and control.

However, neural networks are opaque, limiting our ability to validate them solely from an input-output perspective. This opacity also renders neural network models vulnerable to perturbations \cite{zhang2021adversarial},\cite{yang2022guaranteed}. When it comes to applications in safety-critical scenarios, it requires time-consuming reachability analysis of the specific trajectories for verification, which poses challenges to real-time applications. According to \cite{brix2023first}, the computational efficiency is highly related to the scale of the neural network model. 

This paper aims to promote the interpretability and computational efficiency of neural networks in dynamical system modeling by introducing a novel dual-level modeling framework. Specifically, our proposed approach will divide dynamical system modeling into two essential levels: the low-level neural hybrid system model and its high-level transition system abstraction. The low-level model is employed to precisely capture the system's local behavior and enhance the computational efficiency with a parallel set of shallow neural networks aimed at approximating the local dynamics. Then the high-level transition model, which is an abstraction based on neural hybrid systems, can be obtained based on reachability analysis designed to capture relationships and transition patterns among system subspaces. 
 
The contributions of this paper are summarized as follows.
\begin{itemize}
    \item Maximum Entropy partitioning is applied to partition the system state space into multiple local subspaces, which allows analysis of the dynamics within local subspaces.
    \item A concept of neural hybrid systems is proposed for distributed training and verification of a set of shallow neural networks, thereby enhancing computational efficiency.
    \item A novel transition system abstraction method is proposed to investigate the transition relationships between local partitions, which will further enhance model interpretability.
\end{itemize}

This paper is organized as follows: Preliminaries and problem formulations are given in Section II. The main result, the dual-level modeling framework, is given in Section III. In Section IV, modeling of the LASA data sets is given to illustrate the effectiveness of our proposed framework\footnote{The developed modeling tool and code for experiments are publicly available online at: \url{https://github.com/aicpslab/Dual-Level-Dynamic-System-Modeling}.}. Conclusions are given in Section V.

\subsubsection{Notations.} In the rest of the paper, $\mathbb{N}$ denotes the natural number sets, where $\mathbb{N}^{\le n}$ indicates  $\{1,2,\cdots,n\}$, $\mathbb{R}$ is the field of real numbers, $\mathbb{B}$ is the set of the Boolean variables, $\mathbb{R}^n$ stands for the vector space of $n$-tuples of real numbers, and $\underline{X}$ and $\overline{X}$ are the lower bound and upper bound of an interval $X$, respectively.

\section{Preliminaries and Problem Formulation}

In this paper, the modeling problems for the discrete-time system will be discussed, i.e., we aim to model the system in the form of   
\begin{equation}\label{equ_system dynamics}
    x(k+1)=f(x(k),u(k)),
\end{equation}
in which $x\in \mathbb{R}^{n_x}$ is the system state, $u\in\mathbb{R}^{n_u}$ is the external input, and $f:\mathbb{R}^{n_{x}+n_{u}} \to \mathbb{R}^{n_x}$ is the ideal mapping that precisely describes the system patterns. Due to dimensions and nonlinearity, obtaining $f$ could be challenging, therefore we aim to approximate $f$ with neural network $\Phi:\mathbb{R}^{n_{x}+n_{u}}\to \mathbb{R}^{n_{x}}$ in
\begin{equation}\label{equ_neural dynamic system}
    x(k+1)=\Phi(x(k),u(k)).
\end{equation}

In the training of $\Phi$, approximating $f$ means adjusting the weight and bias of $\Phi$ in order to minimize the error between its output and the given data set. In this paper, the given data set consisting of input-output pairs is in the form of 
\begin{equation}\label{equ_data set}
    \mathcal{D}=\{(z^{(i)},y^{(i)})\mid z^{(i)}\in\mathbb{R}^{n_x+n_u},y^{(i)}\in\mathbb{R}^{n_x}\}.
\end{equation}

However, neural networks face challenges as they typically require extensive data for training and often lack an intuitive understanding of the system's behavior. To unveil the black-box model usually requires a reachability analysis of the neural network dynamical system.

\subsection{Reachability Analysis for Neural Network Dynamical System}

The reachability analysis of neural networks is useful in neural network dynamical system verification for it can determine the range of outputs based on the interplay between the input sets and the structure of the neural network, and according to \cite{tran2019parallelizable,wang2021beta} and \cite{7843595},  simple neural network structure, i.e., $\Phi$ that contains fewer layers and neurons will have advantages in reachable computation. 

Taking a $L$-layer feed-forward neural network $\Phi:\mathbb{R}^{n_0}\to \mathbb{R}^{n_L}$ as an example, its inter-layer propagation can be denoted as follows.
\begin{equation}\label{equ_interpropagation}
    x_{i,k+1} = \sigma(\sum\nolimits_{j} w_{ij,k+1} x_{j,k} + b_{i,k+1}),
\end{equation}
in which $x_{i,k+1}$ is the $i$th neuron output from $k+1$th layer computed by applying the activation function $\sigma$ to the weighted sum of the activations from the previous layer, plus a bias $b_{i,k+1}$ and $w_{ij,k+1}$ is the $i$th line, $j$th row value of the weight bias $W_{k}\in\mathbb{R}^{n_{k}\times n_{k+1}}$. 

Reachability analysis of neural networks will go through the inter-propagation of the neural network in (\ref{equ_interpropagation}), namely, for neural network model in (\ref{equ_neural dynamic system}) output reachable set computation when given $k$th time step state input set $\mathcal{X}_{(k)}\subset\mathbb{R}^{n_x}$ and external input set $\mathcal{U}\subset\mathbb{R}^{n_u}$ can be denoted as 
\begin{equation}\label{equ_reahchable set propagation}
    \mathcal{X}_{(k+1)}=\Phi^*(\mathcal{X}_{(k)},\mathcal{U}),
\end{equation}
in which $\mathcal{X}_{(k+1)}$ is the reachable set output of $\Phi$ at $k+1$th time step computed by reachable set computation method $\Phi^*$ such as \cite{lopez2023nnv,8318388,vincent2021reachable}, etc. Reachable sets in given $K$th time steps requires propagation of (\ref{equ_reahchable set propagation}) in
\begin{equation}
    \mathcal{R}_{(K)}=\bigcup_{k=0}^{K} \mathcal{X}_{(k)},
\end{equation}
in which $\mathcal{R}_{(K)}$ is the reachable sets in $K$time steps. 

Due to the opacity of neural networks, the verification of (\ref{equ_neural dynamic system}) usually necessitates reachability computations of different trajectories to verify specific properties and can be heavily influenced by the neural network structure, posing a computational burden that challenges its application. 

\subsection{Maximum Entropy Partitioning}
Maximum Entropy (ME) partitioning proposed in \cite{10155820} utilizes the Shannon Entropy to partition the state space according to the data, which can be very useful in obtaining subspaces for distributed learning and prediction in neural networks. 

Given a set of $N\in\mathbb{N}$ subspaces $\mathcal{P}=\{\mathcal{P}_i\}_{i=1}^N$, where $\mathcal{P}_i\subset\mathbb{R}^{n_x}$, the Shannon Entropy of $\mathcal{P}$ can be denoted by
\begin{equation}\label{equ_shannon partition}
    H(\mathcal{P})=  -\sum\nolimits_{i=1}^N p(\mathcal{P}_i) \log p(\mathcal{P}_i),
\end{equation}
in which $p({\mathcal{P}})$ denotes the probability of $\mathcal{P}_i$ occurrence in $\mathcal{P}_i$. In this data-driven process, $p(\mathcal{P}_i)$ is extrapolated by the sample set in the form of 

\begin{equation}
    p(\mathcal{P}_i)=\frac{|\mathcal{D}_i|}{|\mathcal{D}|},
\end{equation}
in which $|\mathcal{D}|$ is the number of samples of $\mathcal{D}$ while $D_i$ is defined by
\begin{equation}
    \mathcal{D}_i=\{(z^{(j)},y^{(j)})\in\mathcal{D}\mid x\in\mathcal{P}_i, \forall [x^\top ,u^\top ]^\top =z^{(i)}\}.
\end{equation}

The ME partitioning employs the variation in Shannon entropy from system partitions to ascertain if the current set of partitions maximized the system's entropy after a bisecting method. Explicitly, the variation of Shannon Entropy is in the form of
\begin{equation}
    \Delta H = H(\hat{\mathcal{P}})-H(\mathcal{P})
\end{equation}
in which $\hat{\mathcal{P}}$ is the post-bisecting set of partitions.

By setting a threshold $\epsilon \ge 0$ as a stop condition, namely the bisection process will stop if $\Delta H< \epsilon$, a proper set of partitions can be obtained.

\subsection{Problem Formulation} 

This paper aims to promote the efficiency of learning and prediction of the neural network dynamical system in solving the following problem. 
\begin{problem}\label{problem_1}
    Given the data set $\mathcal{D}$ in the form of (\ref{equ_data set}), how do we model the dynamical system distributively with multiple simple neural networks?
\end{problem}
To promote the interpretability of the learning model, the following problem will be the main concern after a distributed neural network model is obtained.
\begin{problem}\label{problem_2}
    Given a neural-network-based approximation $\Phi$ of $f$, how do we abstract $\Phi$ into an interpretable model that avoids real-time reachable set computation in (\ref{equ_reahchable set propagation})?    
\end{problem}

Solving Problem \ref{problem_1} will allow parallel training and verification of multiple simple neural networks, which enhances the efficiency of neural network modeling while providing an accurate low-level model. Based on the low-level model, we are able to enhance the interpretability by abstracting the low-level model into a high-level model by solving Problem \ref{problem_2}.

\section{Dual-Level Modeling Framework}

Before presenting the dual-level modeling framework, we make the assumption that the system training set (\ref{equ_data set}) provides adequate information in the working zone for dynamical learning as follows.

\begin{assumption}\label{ass_adequte data set}
The working zone of ideal system dynamical description $f$ in (\ref{equ_system dynamics}) is within the localized state space $x\in\mathcal{X}$, given the external input bound where $u\in[\underline{u},\overline{u}]$. 
\end{assumption}

In most cases of neural network dynamical system modeling, $\Phi$ in (\ref{equ_neural dynamic system}) will have high accuracy in approximating the dynamics based on a sample set $\mathcal{D}$. Based on $\mathcal{D}$, we assume that the learning model applies only to a working zone with Assumption \ref{ass_adequte data set}. 

\subsection{Neural Hybrid System Model and Transition System Abstraction}
To solve Problem \ref{problem_1}, we proposed the neural hybrid system model, which allows precise learning of the dynamical system through multiple small-scale neural networks. The neural hybrid system model is defined as 
\begin{definition}\label{def_neural hybrid system}
    A neural hybrid system model is a tuple $\mathcal{H} = \langle \mathcal{P}, \Omega, \delta, \tilde{\Phi} \rangle$ where
\begin{itemize}
    \item $\Omega \subset \mathbb{R}^d$: Working zone, with states $x(k) \in \Omega$.
    \item $\mathcal{P} = \{\mathcal{P}_1, \mathcal{P}_2, \ldots, \mathcal{P}_{N_p}\}$: Finite set of non-overlapping partitions in the working zone, where: 1) $\mathcal{P}_i \subseteq \Omega$; 2) $\bigcup\nolimits _{i=1}^N \mathcal{P}_i=\Omega$; 3) $\mathcal{P}_i\cap\mathcal{P}_j=\emptyset,~ \forall i\neq j$.
    \item $\delta: \Omega \to \{1, 2, \ldots, N_p\}$: Function mapping states to partitions $\delta(x(k)) = i$, implies $x(k) \in \mathcal{P}_i$.
    \item $\tilde{\Phi} = \{\Phi_1, \Phi_2, \ldots, \Phi_{N_p}\}$: Set of neural networks, each $\Phi_i$ models dynamics in $\mathcal{P}_i$.
\end{itemize}
\end{definition}
Definition \ref{def_neural hybrid system} introduces a distributed structure of the neural networks that allows local approximations of the subspaces of state space called partitions. The dynamics of low-level model $\mathcal{H}$ is denoted as
\begin{equation}\label{equ_neural hybrid dynamic system}
    x(k+1)=\Phi_{\delta(x(k))}(x(k),u(k)).
\end{equation}

This distributive structure will help reduce the scales of the neural network approximation and result in the enhancement of the computational efficiency in training and verification. 

Compared with the conventional model, the neural hybrid system modeling will have the advantages of real-time computation and verification. However, to gain insights from the neural hybrid system modeling and enhance interactivity between the learning model and human users, we can abstract the neural hybrid system into a transition system defined as
\begin{definition}\label{def_transition system abstraction}
    A transition system abstraction is a tuple $\mathcal{T} \triangleq \left \langle \Omega,\mathcal{Q},\mathcal{E}\  \right \rangle$ where its elements are:
\begin{itemize}

\item $\Omega\subset \mathbb{R}^{n_x} $: Working zone, where this abstraction is applying to. 

\item $\mathcal{Q}=\{\mathcal{Q}_1,\cdots, \mathcal{Q}_{N_q}\}$: The finite set of subspaces called cells, where: 1) $\mathcal{Q}_i \subseteq \Omega$; 2) $\Omega=\bigcup_{i=1}^{N_q}{\mathcal{Q}_i} $; 3) $\mathcal{Q}_i\bigcap \mathcal{Q}_j=\emptyset$. With an index function $idx:\mathcal{Q}\to\mathbb{N}^{\le N_q}$ for $idx({\mathcal{Q}_i})=i$. 

\item $R:\mathbb{N}^{\le {N_q}}\times \mathbb{N}^{\le N_q}\to\mathbb{B}$: Transition rules, if there exist a probable transition from $\mathcal{Q}_i$ to $\mathcal{Q}_j$, then $R(i,j)=1$, else $R(i,j)=0$.

\end{itemize}
\end{definition}

A transition system abstraction will unveil the interconnection of subspaces with transition rules $T$ through abstracting the neural hybrid system $\mathcal{H}$. In this process, the data in the form of traces will be generated by the neural hybrid system $\mathcal{H}$ by giving it randomized initial states, and randomized or user-specified external input for the non-autonomous dynamical systems.  

\subsection{Efficient Dynamics Learning via Low-Level Modeling}
In this paper, we will be achieving efficient dynamics learning via our proposed low-level model, namely, neural hybrid system modeling. To begin with, ME-partitioning proposed in \cite{10155820} will be applied to bisecting the working zone $\Omega$ based on the data set $\mathcal{D}$. In this process, $\Omega$ and $\mathcal{P}$ will be in the form of the interval, e.g., $\mathcal{P}_i=[\underline{p}_{i,1},\overline{p}_{i,1}]\times[\underline{p}_{i,2}\times\overline{p}_{i,2}]\ldots$ in which $\overline{\mathcal{P}}_i=\{\overline{p}_{i,1},\overline{p}_{i,2},\cdots,\overline{p}_{i,n_x}\}\in\mathbb{R}^{n_x},$ etc. Specifically, we will locate the $j$th dimension of the $i$th partition to bisect via
\begin{equation}\label{equ_find distance}
    ({i,j})=\arg\max\nolimits_{i,j} D_{i,j},
\end{equation}
in which
\begin{equation}
    D_{i,j}=\overline{p}_{i,j}-\underline{p}_{i,j}.
\end{equation}
 We will keep bisecting the $\mathcal{P}$ until $\Delta H \le \epsilon$. After the ME partitioning, the set of partitions $\mathcal{P}$ with $N_p$ partitions can be obtained, which will subsequently define the segmented data set $\{\mathcal{D}_1,\ldots,\mathcal{D}_{N_p}\}$. With the segmented data set, we are able to train the set of neural networks once given a neural network structure, namely, the layers, neurons, and the activation function, etc., of neural networks.


To further optimize the ME partitioning and simplify the learning model, we will merge the redundant partitions based on the training performance of the neural network. Merging redundant partitions will be based on the Mean Square Error (MSE) performance of the neural network. Given $\mathcal{D}$ and a trained neural network $\Phi$, the MSE performance of $\Phi$ is 
\begin{equation}
    MSE(\Phi,\mathcal{D})=\frac{1}{|\mathcal{D}|}\sum\nolimits_{i=1}^{|\mathcal{D}|}\left\| \Phi(z^{(i)})-y^{(i)}\right\|.
\end{equation}

By setting a threshold based on MSE performance $\gamma \ge 0$, we are able to identify the redundant partitions that are considered to have similar performance under the same neural network structure, namely, if
\begin{equation}
    MSE(\Phi,\mathcal{D}_i \cup \mathcal{D}_j)\le\gamma
\end{equation}
for a trained $\Phi$, the corresponding partitions $\mathcal{P}_i$ and $\mathcal{P}_j$ will be considered redundant partitions, and hence they will be merged. 

Merging the redundant partitions will subsequently define the switching logic $\delta$ and the set of neural networks $\tilde{\Phi}$ for the neural hybrid system $\mathcal{H}$.  The low-level neural hybrid system modeling can be summarized in pseudo-code given in the Algorithm \ref{alg:combined_example}.

\begin{algorithm}
\caption{Low-Level Neural Hybrid System Modeling}
\label{alg:combined_example}
\begin{algorithmic}[1] 
\Statex \Comment{Maximum Entropy partitioning}
\Procedure{ME Partitioning}{$\Omega,\mathcal{D},\epsilon$} 
    \Statex \textbf{Input:} $\Omega;\mathcal{D};\epsilon$.
    \Statex \textbf{Output:} $\mathcal{P}$; $\cup\{\mathcal{D}_i\}$.
   \State $P_{save} \gets \emptyset$; $\mathcal{D}_{save} \gets \emptyset$;
\State $P_1 \gets \Omega$;
\While{$\exists \Delta H_i \ge \epsilon,~\forall \mathcal{P}_i$ }
    \State $[i,j,Distance] \gets \max(D_{i,j})$
    \State Obtain $\mathcal{P}_{temp1}$ and $\mathcal{P}_{temp2}$ under (\ref{equ_find distance})
    \State Obtain $\mathcal{D}_{temp1}$ and $\mathcal{D}_{temp2}$ 
    \If{$\Delta H_i \ge entropy$} \Comment{Using (\ref{equ_shannon partition})}
        \State $P_i \gets \{P_{temp1}, P_{temp2}\}$
        \State $\mathcal{D}_{i}\gets \{\mathcal{D}_{temp1}, \mathcal{D}_{temp2}\}$
    \Else
        \State Add $P_{i}$ to $P_{save}$ and delete $P_i$
        \State Add $\mathcal{D}_{i}$ to $\mathcal{D}_{save}$ and delete $\mathcal{D}_i$
    \EndIf
\EndWhile
    \State \textbf{return} $\mathcal{P}\cup \mathcal{P}_{save}$; $\mathcal{D}\cup\mathcal{D}_{save}$.
\EndProcedure

\Statex \Comment{Merging and dynamics learning}
\Procedure{Merge and Learn}{$\mathcal{P},\cup\{\mathcal{D}_i\},\Phi$} 
    \Statex \textbf{Input:} $\mathcal{P},\cup\{\mathcal{D}_i\},\Phi$
    \Statex \textbf{Output:} $\mathcal{P},\tilde{\Phi}$
    \State $\ell \gets |\mathcal{P}|$, $N \gets 1$;
\Comment{Segmented partitions merge}
\While{$N < \ell$}
    \State $n \gets 1$;
    \While{$n \le \ell$}
        \State $n \gets n + 1$;
        \State $\Phi_{N,n}\gets\Phi$, $\mathcal{D}_{N,n}\gets{\mathcal{D}_N}\cup\mathcal{D}_n$
        \State $\Phi_{N,n}\gets \arg\min\nolimits_{\Phi_{N.n}}MSE(\Phi_{N,n},\mathcal{D}_{N,n})$
        \If{$MSE(\Phi_{N,n},\mathcal{D}_{N,n}) \le \gamma$}
            \State $\mathcal{P}_{N} \gets \{\mathcal{P}_{N}\cup\mathcal{P}_{n}\}$
            \State Delete $\mathcal{P}_n$
            \State $\ell \gets \ell - 1$
        \EndIf
    \EndWhile
    \State $N \gets N + 1$;
\EndWhile\\
\Comment{Generate neural network approximations}
\State $i \gets 1$;
\While{$i \le N$}
    \State $\Phi_{i}\gets \arg\min\nolimits_{\Phi_{i}}MSE(\Phi_{i},\mathcal{D}_{i})$
\EndWhile\\
\Return $\mathcal{P} = \{\mathcal{P}_1\ldots,\mathcal{P}_N\}$; $\Phi = \{\Phi_1,\ldots,\Phi_N\}$

\EndProcedure
\end{algorithmic}
\end{algorithm}

The low-level neural hybrid system can model the dynamical system through a distributive and computationally efficient framework, which makes it possible for parallel training in the Merge and Learning procedure, and distributive verification in \cite{WANG2023126879}. To further exploit this distributive structure and promote the interpretability of the low-level learning model, we proposed a transition system abstraction method as the high-level model.   

\subsection{Interpretable Abstraction via High-Level Model}
In high-level model abstraction, we intend to abstract the neural hybrid system model in Definition \ref{def_neural hybrid system} into a transition system in Definition \ref{def_transition system abstraction} with the help of the data generated by $\mathcal{H}$ called the set of samples, defined by

\begin{definition}\label{def_abstraction samples}
Set of samples $\mathcal{W}=\{w_1,w_2,\cdots,w_L\}$ of neural hybrid system (\ref{equ_neural hybrid dynamic system}) is a collection of sampled $L$ traces obtained by given $\mathcal{H}$ different initial condition and randomized external input $u\in\mathcal{U}$, where for each trace $w_i$, $i = 1,\ldots,L$, is a finite sequence of time steps and data $(k_{0,i},d_{0,i}),(k_{1,i},d_{1,i}),\cdots,(k_{M_i,i},d_{M_i,i})$ in which
\begin{itemize}
    \item $k_{0,i}\in(0,\infty)$ and $k_{\ell+1,i}=k_{\ell,i}+1$, $\forall \ell\in\mathbb{N}^{\le M_i},\ \forall i\in\mathbb{N}^{\le L}$.
    \item $z_{\ell,i}=[x^\top_{i}(k_{\ell,i}),~u^\top_{i}(k_{\ell,i})]^\top \in \mathbb{R}^{n_x+n_u}$,
    $\forall \ell = 0,1,\ldots,M_i$, $\forall i\in\mathbb{N}^{\le L}$, where $x_{i}(k_{\ell,i}),u_{i}(k_{\ell,i})$ denote the state and input of the system at $\ell$th step for $i$th trace, respectively.
\end{itemize}
\end{definition}
\begin{remark}
    It should be noted that the abstraction of the neural hybrid system is specific, meaning that different transition system abstractions can be obtained based on different control strategies. This specificity aids system designers in implementing and validating control strategies tailored to specific partitions.
\end{remark}

After obtaining the set of samples, the set of cells $\mathcal{Q}$ will be obtained via the ME partitioning method as in procedure ME partitioning in Algorithm \ref{alg:combined_example} based on $\mathcal{W}$. Then, the transition relationships between cells will be computed via reachability analysis in
\begin{equation}\label{equ_hybrid next set estimation}
    \mathcal{Q}'_i=\bigcup\nolimits_{j=1}^{N_p}\Phi_j^* (\mathcal{Q}_i\cap\mathcal{P}_j,\mathcal{U}),
\end{equation}
in which $\Phi_j^*$ indicates a reachable set computation method using the sub-neural network. Intuitively, based on Definition \ref{def_transition system abstraction} the transition rule $R(i,j)$ is 
\begin{equation}
    R(i,j)=
    \begin{cases}
            1,~\mathcal{Q}'_i\cap\mathcal{Q}_j\neq\emptyset\\
            0,~\mathcal{Q}'_i\cap\mathcal{Q}_j=\emptyset
        \end{cases}
\end{equation}
The process of transition computation can be summarized in pseudo-code given in Algorithm \ref{alg2}.
\begin{algorithm}[t!]
\caption{Transition Computation via $\mathcal{H}$}
\label{alg2}
\begin{algorithmic}[1]
\Statex \textbf{Input:} $\mathcal{P},\mathcal{Q},\Phi,\mathcal{U}$
\Statex \textbf{Output:} $R$
\State  $N_p\gets|\mathcal{P}|$, $N_q\gets|\mathcal{Q}|$
\State $i \gets 1$; $j \gets 1$;
\While{$i \le n$}
    \State $\mathcal{Q}_i'\gets\bigcup_{l=1}^{N_p}\Phi^*_l(\mathcal{Q}_i\cap P_l,\mathcal{U})$ \Comment{  Using (\ref{equ_hybrid next set estimation})} 
    \State $j \gets 1$
    \While{$j \le n$}
        \If{$\mathcal{Q}'_i \cap \mathcal{Q}_j \neq \emptyset$}
            \State $R(i,j) \gets 1$
        \Else
            \State $R(i,j) \gets 0$
        \EndIf
        \State $j \gets j + 1$
    \EndWhile
    \State $i \gets i + 1$
\EndWhile\\
\Return $R$
\end{algorithmic}
\end{algorithm}

The proposed dual-level modeling framework can be summarized as follows.
\begin{itemize}
    \item The localized working zone of $\Omega$, i.e., $\mathcal{P}$ can be obtained based on an ME partitioning process, which is completely data-driven and can be easily tuned by adjusting the threshold.  
    \item Partitions can be further optimized based on the MSE performance of the trained neural network to simplify the low-level model.
    \item The low-level model has a distributive structure consisting of simple neural networks that allow parallel training and verification, which will be computationally efficient. 
    \item The low-level model can be further abstracted into a high-level transition system, this process can be specifically designed and allow system designers to develop and test control strategies that are specifically tailored for each distinct localized cell.
    \item The transitions can be off-line computed by reachability analysis, and can be transferred into a transition graph which will enhance the learning model's interpretability, and enable the feasibility of verifications based on logical descriptions.
\end{itemize}

\section{Applications to Dynamical System Modeling}

Regarding the modeling of complex dynamical systems like human behaviors, learning-based methodologies have garnered significant attention for their efficacy in \cite{reinhart2011neural,kanazawa2019learning}, etc. However, while learning-based approaches offer advantages over mechanistic modeling, they present numerous challenges in practical applications. For instance, typical issues include:
\begin{itemize}
    \item The limited availability of sample data may result in a deep neural network-based dynamical system model that is not adequately trained, thereby hindering its ability to capture the full spectrum of human behavioral complexities.
    \item The inherent nature of human demonstrations, characterized by sudden shifts, suggests that a trained neural network-based dynamical system model might exhibit discrepancies in its behavior, especially in localized regions of the operational space.
    \item The interpretability deficit in neural network-based dynamical system models poses a significant challenge in real-time applications for limited, and computationally intensive verification methods.
\end{itemize}

\begin{figure}[ht]
\centering
	\subfloat[$MultiModels_2$ Handwriting Human Demonstration and $37$ Partitions Obtained.]{\includegraphics[width=0.25\textwidth]{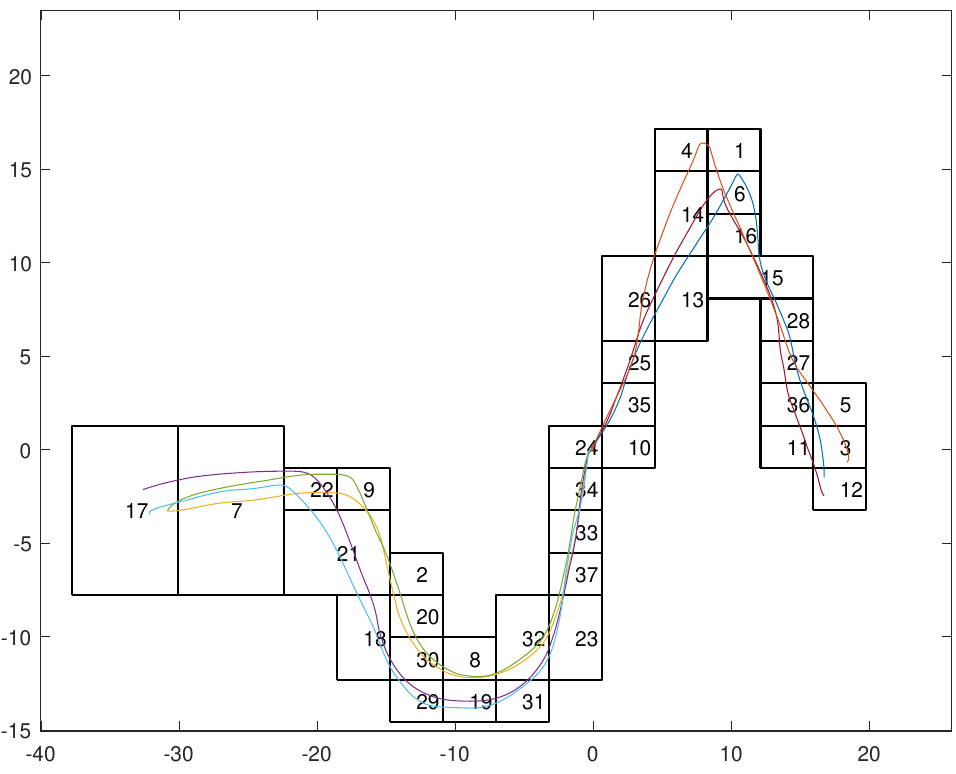}}
    \subfloat[$12$ partitions obtained where redundant ones are given in the same color.]{\includegraphics[width=0.25\textwidth]{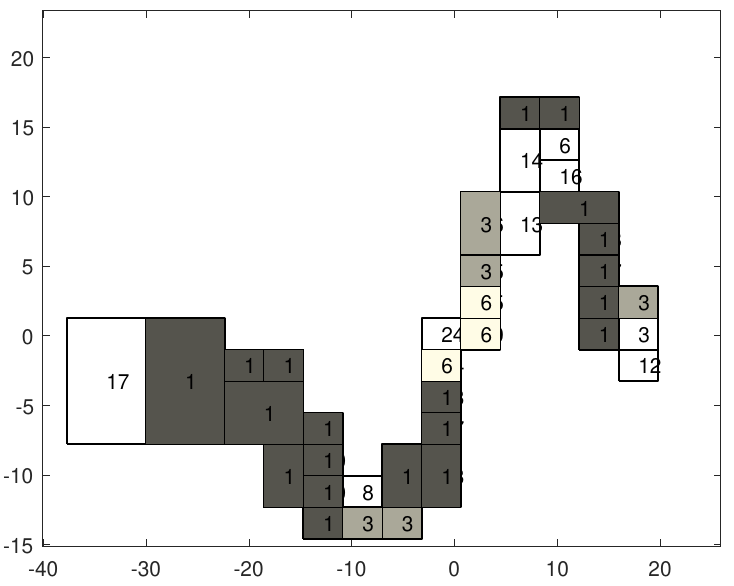}} \par
	\subfloat[$11$ Cells Abstraction for $\mathcal{H}$ of $MultiModels_2$.]{\includegraphics[width=0.25\textwidth]{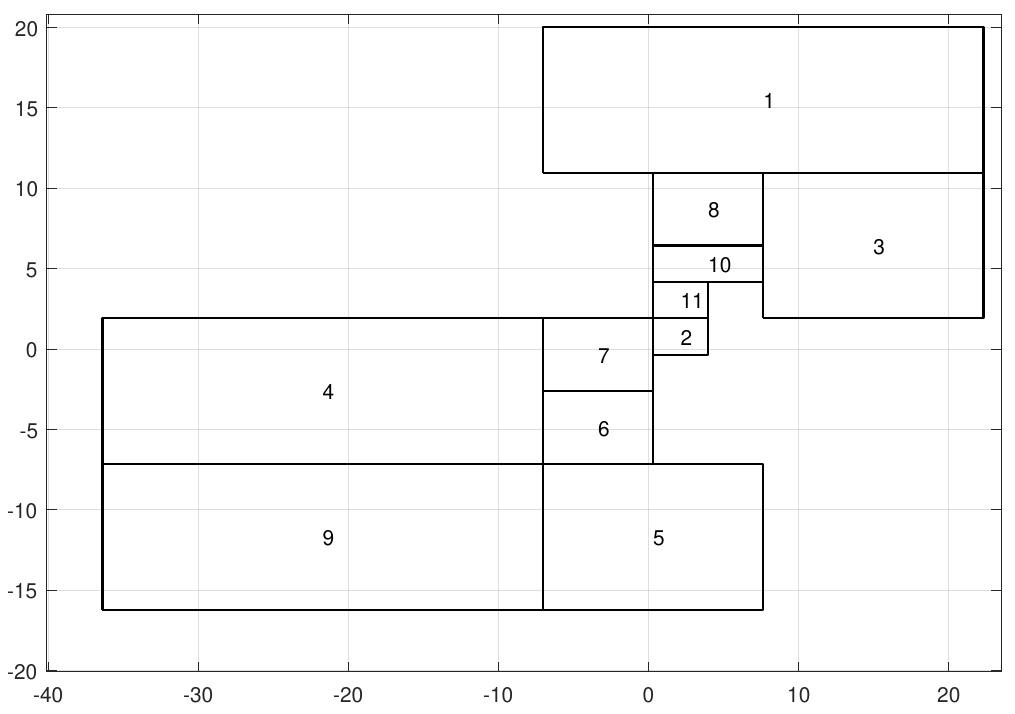}}
	\subfloat[Transition Map based on $\mathcal{T}$ of $MultiModels_2$.]{\includegraphics[width=0.25\textwidth]{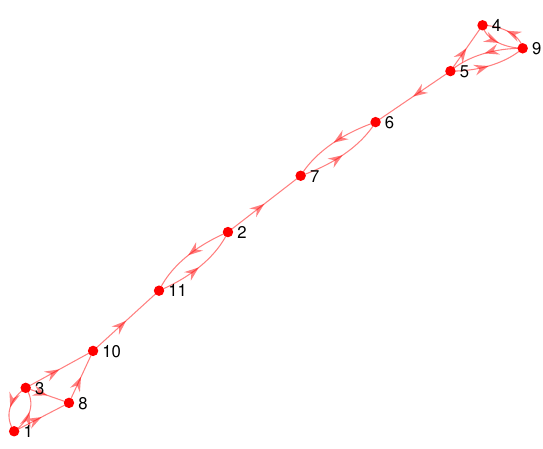}}
	\caption{Partitions, Cells, and Transition Map Abstraction of Dual-Level Models for $MultiModels_2$ from LASA data set.}
	\label{Fig_LASA} 
\end{figure}

The above issues are exemplified in the LASA dataset \cite{5953529} modeling, which encompasses a diverse range of handwriting motions demonstrated by human users across 30 distinct shapes. This paper will attempt to address these issues through our proposed dual-level dynamical system modeling framework. The dual-level modeling process can be summarized as follows.
\begin{itemize}
   \item Extreme Learning Machines (ELMs) are employed, each comprising 20 ReLU-activated neurons. These ELMs feature a randomized input weight matrix and bias vector, forming the core structure of the model. To highlight the efficacy of our modeling approach, an ELM with a solitary hidden layer containing 200 ReLU-activated neurons is trained to serve as a single-neural network reference model.
   \item A threshold of $\epsilon=4\times 10^{-2}$ is set for ME partitioning variation in Algorithm \ref{alg:combined_example}. This setting led to the generation of the set of partitions in ME partitioning for the low-level model of all $30$ shapes some results are given in Tab. \ref{tb:shape_training1} and Tab. \ref{tb:shape_reference2}, an illustration of $MutiModels_2$\footnote{Complete results includes modeling for all $30$ shapes can be found on our GitHub repository on dual-level dynamical system modeling at \url{https://github.com/aicpslab/Dual-Level-Dynamic-System-Modeling/tree/main/Results}} is given in Fig. \ref{Fig_LASA} (a). 
   \item By setting a threshold $\gamma=1.5\times 10^{-5}$ in Merging, we manage to simplify the low-level model by training fewer neural networks while maintaining accuracy.
   \item We obtain the abstraction data from randomly generated trajectories in the working zone $\Omega$, where $\forall M_i=400,~\forall i\in\mathbb{N}^{\le 400}$ under Definition \ref{def_abstraction samples}. By applying the threshold $\epsilon=4\times 10^{-2}$, a set of cells is then generated, as shown in Fig. \ref{Fig_LASA} (c).
   \item Based on the set of cells, we employ the Algorithm \ref{alg2} to compute the transition relationships of the high-level model abstraction. The transition from Fig. \ref{Fig_LASA} (c) to Fig. \ref{Fig_LASA} (d) allows for the interpretation of transition relationships between local working zones.
   \item We verify the transition system abstraction via Computation Tree Logic (CTL) formulae \cite{pan2016model}, in which $\diamond$ or $\Box$ denote the for $some$ or $all$ traces, and $\bigcirc$ denotes the next step. The formulae and results of $MultiModel_2$ are given in Tab. \ref{pro_1_tab} as examples, $\phi_1$ indicates the possibility of the neural hybrid system model being in $\mathcal{Q}_2$, $\phi_2$ specifies that for every possible next step, the system will be in $\mathcal{Q}_4$, and $\phi_3$ signifies whether there exists a trajectory that reaches $\mathcal{Q}_7$ immediately after passing through $\mathcal{Q}_6$, given the initial condition is $\mathcal{Q}_9$.  
   
\end{itemize}

\begin{table}[hb]\begin{center}\caption{Training Time and MSE of the Low-level Model}\label{tb:shape_training1}\begin{tabular}{|ccc|}\hline Shape Name & Training Time (ms) & MSE ($10^{-5}$) \\ \hline $Khamesh$ & $0.7147$ & $0.3466$ \\  $LShape$ & $0.7784$ & $0.3745$ \\   $Multi Models_1$ & $0.5603$ & $0.2858$ \\  $Multi Models_2$ & $0.6225$ & $0.2892$ \\$\cdots$ & $\cdots$ & $\cdots$  \\  \hline \end{tabular}\end{center}\end{table}

\begin{table}[hb]\begin{center}\caption{Training Time and MSE for ELM Model}\label{tb:shape_reference2}\begin{tabular}{|ccc|}\hline Shape Name & Training Time (ms) & MSE ($10^{-5}$) \\ \hline $Khamesh$ & $14.8098$ & $0.0923$ \\  $LShape$ & $38.4118$ & $0.0842$ \\  $Multi Models_1$ & $20.8079$ & $0.0551$ \\  $Multi Models_2$ & $26.6117$ & $0.0753$ \\  $\cdots$ & $\cdots$ & $\cdots$  \\  \hline \end{tabular}\end{center}\end{table}

\begin{table}[t!]
	\centering
	\caption{Verification results of $CTL$ formula:  $\mathcal{T}_{MultiModels_2}$ with $\mathcal{Q}_{9}$ as the initial cell. }\label{pro_1_tab}
	\label{tab5}
	\begin{tabular}{|cc|}
		\hline 
        \textbf{$CTL$ formula}  &\textbf{$\mathcal{T}_{MultiModels_2}$} \\
		\hline\hline
$\phi_1=\exists\diamond \mathcal{Q}_{2}$   &$true$  \\
$\phi_2=\forall \bigcirc \mathcal{Q}_{4}$   &$false$  \\
$\phi_3=\exists( \mathcal{Q}_6) \wedge (\exists\bigcirc\mathcal{Q}_{7})$   &$true$  \\
\hline
	\end{tabular}
\end{table}

\section{Conclusion}
In this paper, a dual-level dynamical system learning framework is proposed to promote computational efficiency and interpretability in system identification. This framework utilizes a data-driven ME partitioning process to bisect the working zone, which makes it possible for parallel training and local analysis. In order to simplify the learning model, a process called Merging is proposed to merge the partitions based on the training performance. The low-level model is then able to learn the dynamics precisely while only consisting of a set of simple neural networks. A high-level model is proposed to promote interpretability through the reachability analysis. This high-level model will provide valuable insights into the transition relationship within the working zone with the transition map and allow user-specified verification through CTL formulae.

\bibliography{ifacconf}             

\begin{thebibliography}{17}
\providecommand{\natexlab}[1]{#1}
\providecommand{\url}[1]{\texttt{#1}}
\providecommand{\urlprefix}{URL }
\expandafter\ifx\csname urlstyle\endcsname\relax
  \providecommand{\doi}[1]{doi:\discretionary{}{}{}#1}\else
  \providecommand{\doi}{doi:\discretionary{}{}{}\begingroup \urlstyle{rm}\Url}\fi

\bibitem[{Brix et~al.(2023)Brix, M{\"u}ller, Bak, Johnson, and Liu}]{brix2023first}
Brix, C., M{\"u}ller, M.N., Bak, S., Johnson, T.T., and Liu, C. (2023).
\newblock First three years of the international verification of neural networks competition (vnn-comp).
\newblock \emph{International Journal on Software Tools for Technology Transfer}, 1--11.

\bibitem[{Feng et~al.(2018)Feng, Leung, and Sum}]{7843595}
Feng, R., Leung, C.S., and Sum, J. (2018).
\newblock Robustness analysis on dual neural network-based $k$ wta with input noise.
\newblock \emph{IEEE Transactions on Neural Networks and Learning Systems}, 29(4), 1082--1094.

\bibitem[{Kanazawa et~al.(2019)Kanazawa, Zhang, Felsen, and Malik}]{kanazawa2019learning}
Kanazawa, A., Zhang, J.Y., Felsen, P., and Malik, J. (2019).
\newblock Learning 3d human dynamics from video.
\newblock In \emph{Proceedings of the IEEE/CVF conference on computer vision and pattern recognition}, 5614--5623.

\bibitem[{Khansari-Zadeh and Billard(2011)}]{5953529}
Khansari-Zadeh, S.M. and Billard, A. (2011).
\newblock Learning stable nonlinear dynamical systems with gaussian mixture models.
\newblock \emph{IEEE Transactions on Robotics}, 27(5), 943--957.

\bibitem[{Lopez et~al.(2023)Lopez, Choi, Tran, and Johnson}]{lopez2023nnv}
Lopez, D.M., Choi, S.W., Tran, H.D., and Johnson, T.T. (2023).
\newblock Nnv 2.0: The neural network verification tool.
\newblock In \emph{International Conference on Computer Aided Verification}, 397--412. Springer.

\bibitem[{Pan et~al.(2016)Pan, Li, Cao, and Ma}]{pan2016model}
Pan, H., Li, Y., Cao, Y., and Ma, Z. (2016).
\newblock Model checking computation tree logic over finite lattices.
\newblock \emph{Theoretical computer science}, 612, 45--62.

\bibitem[{Reinhart and Steil(2011)}]{reinhart2011neural}
Reinhart, R.F. and Steil, J.J. (2011).
\newblock Neural learning and dynamical selection of redundant solutions for inverse kinematic control.
\newblock In \emph{2011 11th IEEE-RAS International Conference on Humanoid Robots}, 564--569. IEEE.

\bibitem[{Stefenon et~al.(2022)Stefenon, Corso, Nied, Perez, Yow, Gonzalez, and Leithardt}]{stefenon2022classification}
Stefenon, S.F., Corso, M.P., Nied, A., Perez, F.L., Yow, K.C., Gonzalez, G.V., and Leithardt, V.R.Q. (2022).
\newblock Classification of insulators using neural network based on computer vision.
\newblock \emph{IET Generation, Transmission \& Distribution}, 16(6), 1096--1107.

\bibitem[{Tran et~al.(2019)Tran, Musau, Lopez, Yang, Nguyen, Xiang, and Johnson}]{tran2019parallelizable}
Tran, H.D., Musau, P., Lopez, D.M., Yang, X., Nguyen, L.V., Xiang, W., and Johnson, T.T. (2019).
\newblock Parallelizable reachability analysis algorithms for feed-forward neural networks.
\newblock In \emph{2019 IEEE/ACM 7th International Conference on Formal Methods in Software Engineering (FormaliSE)}, 51--60. IEEE.

\bibitem[{Vincent and Schwager(2021)}]{vincent2021reachable}
Vincent, J.A. and Schwager, M. (2021).
\newblock Reachable polyhedral marching (rpm): A safety verification algorithm for robotic systems with deep neural network components.
\newblock In \emph{2021 IEEE International Conference on Robotics and Automation (ICRA)}, 9029--9035. IEEE.

\bibitem[{Wang et~al.(2021)Wang, Zhang, Xu, Lin, Jana, Hsieh, and Kolter}]{wang2021beta}
Wang, S., Zhang, H., Xu, K., Lin, X., Jana, S., Hsieh, C.J., and Kolter, J.Z. (2021).
\newblock Beta-crown: Efficient bound propagation with per-neuron split constraints for neural network robustness verification.
\newblock \emph{Advances in Neural Information Processing Systems}, 34, 29909--29921.

\bibitem[{Wang et~al.(2023{\natexlab{a}})Wang, Yang, and Xiang}]{WANG2023126879}
Wang, T., Yang, Y., and Xiang, W. (2023{\natexlab{a}}).
\newblock Computationally efficient neural hybrid automaton framework for learning complex dynamics.
\newblock \emph{Neurocomputing}, 562, 126879.

\bibitem[{Wang et~al.(2023{\natexlab{b}})Wang, Wang, Chen, Huang, Nguyen, De, and Hussain}]{wang2023fusing}
Wang, Y., Wang, W., Chen, Q., Huang, K., Nguyen, A., De, S., and Hussain, A. (2023{\natexlab{b}}).
\newblock Fusing external knowledge resources for natural language understanding techniques: A survey.
\newblock \emph{Information Fusion}, 92, 190--204.

\bibitem[{Xiang et~al.(2018)Xiang, Tran, and Johnson}]{8318388}
Xiang, W., Tran, H.D., and Johnson, T.T. (2018).
\newblock Output reachable set estimation and verification for multilayer neural networks.
\newblock \emph{IEEE Transactions on Neural Networks and Learning Systems}, 29(11), 5777--5783.

\bibitem[{Yang et~al.(2022)Yang, Wang, Woolard, and Xiang}]{yang2022guaranteed}
Yang, Y., Wang, T., Woolard, J.P., and Xiang, W. (2022).
\newblock Guaranteed approximation error estimation of neural networks and model modification.
\newblock \emph{Neural Networks}, 151, 61--69.

\bibitem[{Yang and Xiang(2023)}]{10155820}
Yang, Y. and Xiang, W. (2023).
\newblock Modeling dynamical systems with neural hybrid system framework via maximum entropy approach.
\newblock In \emph{2023 American Control Conference (ACC)}, 3907--3912.

\bibitem[{Zhang et~al.(2021)Zhang, Zheng, and Mao}]{zhang2021adversarial}
Zhang, X., Zheng, X., and Mao, W. (2021).
\newblock Adversarial perturbation defense on deep neural networks.
\newblock \emph{ACM Computing Surveys (CSUR)}, 54(8), 1--36.

\end{thebibliography}
                                                   







\end{document}